\documentclass[12pt,a4paper]{iopart}
\usepackage{bbm}
\usepackage{graphicx}
\usepackage{oldgerm}
\usepackage{mathpak}
\usepackage{amssymb}

\newcommand{\bra}[1]{\ensuremath{\left\langle #1\right|}}
\newcommand{\ket}[1]{\ensuremath{\left|#1\right\rangle}}

\begin{document}

\title{\mbox{A geometric comparison of entanglement}\\ \mbox{and quantum nonlocality in discrete systems}}

\author{Christoph Spengler, Marcus Huber and Beatrix C. Hiesmayr}
\address{Faculty of Physics, University of Vienna, Boltzmanngasse 5, 1090 Vienna, Austria}
\ead{\mailto{Christoph.Spengler@univie.ac.at}, \mailto{Marcus.Huber@univie.ac.at}, \mailto{Beatrix.Hiesmayr@univie.ac.at}}

\begin{abstract}
We compare entanglement with quantum nonlocality employing a geometric structure of the state space of bipartite qudits. Central object is a regular simplex spanned by generalized Bell states. The Collins--Gisin--Linden--Massar--Popescu--Bell inequality is used to reveal states of this set that cannot be described by local-realistic theories. Optimal measurement settings necessary to ascertain nonlocality are determined by means of a recently proposed parameterization of the unitary group $\mathcal{U}(d)$ combined with robust numerical methods. The main results of this paper are descriptive geometric illustrations of the state space that emphasize the difference between entanglement and quantum nonlocality. Namely, it is found that the shape of the boundaries of separability and Bell inequality violation are essentially different. Moreover, it is shown that also for mixtures of states sharing the same amount of entanglement, Bell inequality violations and entanglement measures are non-monotonically related.
\end{abstract}

\maketitle

\newpage

\section{Introduction}
The fact that quantum physics contradicts local-realism is one of
the most seminal discoveries. In his pioneering work \cite{BellORIGINAL}, John S. Bell showed that the statistical behavior of a
bipartite qubit system is irreconcilable with any local-realistic
theory when the system is in a singlet state. He
revealed that for such theories correlations are bounded by constraints
which, however, can be violated if the setup is handled
quantum physically. Since then much effort has been made to fully
understand the origins and conditions that allow us to demonstrate and experimentally test this
contradiction known as quantum nonlocality. Nowadays, a variety of so-called Bell
inequalities is known. Recent Bell inequalities also cover systems with more than two degrees of freedom \cite{CGLMPineq} and/or more than two parties
\cite{Mermin,WWZB1,WWZB2,svozil,threequtrits}.

Yet, the relation between entanglement
and nonlocality is not completely clarified apart from the fact that Bell inequality violations are a consequence of the presence of entanglement \cite{Terhal,Verstraete,Hyllus}. The existence of entangled states that do not violate any Bell
inequality \cite{Werner} has raised several questions. First, how far can Bell inequalities be improved and is it possible for a given system to systematically derive the most powerful inequality or a set of
those (see \cite{Davis,Navascues1,Navascues2,Bancal} for recent approaches)? Second, which entangled states allow a
local--realistic description even if LOCC (local operators and
classical communication) and POVMs (positive operator
valued measurements) are used and what are their characteristics? Besides that, a connection between
the violation of a Bell inequality and the security of quantum cryptography was shown, attesting to the importance of further investigations \cite{security1,security2}.

A nontrivial connection between nonlocality and entanglement was found by investigating the Collins--Gisin--Linden--Massar--Popescu--Bell (CGLMP) inequality \cite{CGLMPineq}. In particular, it was shown that maximal violation is achieved with non-maximally entangled states
\cite{AcinDurt,Junge}. Similar features were also found in \cite{H2} for neutral K--mesons, which are two level
systems oscillating and decaying in time. For this system entangled in strangeness, though having only two degrees of freedom one finds no violation of the CHSH--Bell inequality if the system is prepared in the spin singlet state copiously produced in accelerator experiments \cite{DiDomenico}. A violation is only achieved if a non--maximally entangled state is used. Furthermore, a relation between Bell inequality violation and the violation of the $\mathcal{CP}$ symmetry ($\mathcal{C}$ - charge conjugation, $\mathcal{P}$ - parity) in particle physics was established \cite{BGH3,BH1}. (It should be noted that this particular system in high energy physics is considerably different from systems not being affected by decay processes).  Another example demonstrating the fundamental difference between entanglement and nonlocality is given in \cite{nonlocalmachine}. Here, it is shown that the simulation of non-maximally entangled states via so-called non-local boxes (or Popescu-Rohrlich boxes) requires more
resources than the simulation of a Bell state.

In the present work we aim to study nonlocality and entanglement in the context of state space geometry. That is, we consider these properties using descriptive real vector representations for density matrices. Our main motivation is the fact that geometric considerations of the state space can provide deeper insights into quantum physics. They are also ideal to obtain an impression of the volumes related to certain properties or the strength of particular criteria. In case of a single qubit a geometric representation is given by the well-known three dimensional Bloch-vector where the state space is called the Bloch-sphere. Regarding multipartite systems one usually tries to find representations where local properties and correlations can be separated in different quantities in order to avoid unwanted complexity caused by high-dimensionality. Hence, when it comes to studying entanglement and nonlocality of a certain system it suffices to investigate representatives of equivalence classes of states which are equivalent with respect to local unitary transformations. In this way, for bipartite qubits it was found that correlations can be summarized in a three--dimensional vector lying within a regular tetrahedron \cite{HorodeckiGEOMETRY,BertlmannGEOMETRY}. For the investigation of bipartite qudit systems we utilize a proposed generalization of this tetrahedron -- a regular simplex spanned by mutually orthogonal maximally entangled states \cite{BHNsimplexqutrits,BHNsimplex,BHNsimplexqutrits2}. Our goal is to visualize and compare the geometries of the boundaries of entanglement and nonlocality of low-dimensional subsections of this set of states.

In order to do so, we choose certain classes of states for which via optimal entanglement
witnesses we are able to exactly determine the boundary between separable and entangled states. For the same classes we then reveal all states that violate the CGLMP--Bell inequality. Determining, for a given state, whether it violates or obeys a certain Bell inequality is a high--dimensional nonlinear optimization
problem which is difficult to solve even for low--dimensional quantum systems. In Sec.~\ref{nonlocalstates} we show how this problem can be solved utilizing an advantageous parameterization of the unitary group $\mathcal{U}(d)$ and robust numerical methods.

Our main results concerning the geometry of nonlocality and entanglement are given in Sec.~\ref{visCHSH} for bipartite qubits and in Sec.~\ref{visqutrits} for bipartite qutrits. In Sec.~\ref{measure} we extend our geometric considerations by incorporating a comparison between Bell inequality violation and an entanglement measure. Our results show that both quantities are so intrinsically different that not even for mixtures of states sharing the same amount of entanglement the relation is monotone.

\section{Foundations of Bell inequalities}
We begin by reviewing the foundations of Bell
inequalities. Any local-realistic description of a bipartite system
with $d$ outcomes and $n$ measurement apparatuses on each side
implies the existence of $d^{2n}$ probabilities
\begin{align}
\label{probdist}
P(A_1=j,\ldots,A_n=k;B_1=l,\ldots,B_n=m)\geq0
\end{align}
with normalization
\begin{align}
\label{norm}
\sum_{j,\ldots,m=0}^{d-1} P(A_1=j,\ldots,A_n=k;B_1=l,\ldots,B_n=m)=1
\end{align}
determining the statistics of the observables
$A_1,\ldots,A_n,B_1,\ldots,B_n$ which can take on the values
$j,\ldots,k,l,\ldots,m \in \{0,\dots,d-1\}$
\cite{Davis,PeresATBI,Masanes}. As it is known, probability
distributions of this form cannot reproduce the statistical behavior
of composite quantum systems
$\mathcal{H}^{AB}=\mathbb{C}^d\otimes\mathbb{C}^d$ when the system
is in a certain state and the observables are chosen appropriately. In
order to reveal this, one constructs quantities $I$ that
consist of linear combinations of joint probabilities
\begin{align*}
&P(A_a=x;B_b=y)=\\
\sum_{j,\ldots,m=0\ (except \ x,y)}^{d-1} &P(A_1=j,\ldots,A_a=x,\ldots,A_n=k;B_1=l,\ldots,B_b=y,\ldots,B_n=m)
\end{align*}
and shows that they are bounded, i.e.
\begin{align}
\label{Bellineq}
I\leq \Omega
\end{align}
if (\ref{probdist}) and (\ref{norm}) are assumed. If this bound can
be exceeded when the system is treated quantum physically, then the
established inequality (\ref{Bellineq}) is called a Bell
inequality.

\section{Determining nonlocal quantum states}
\label{nonlocalstates}

States that cannot be described by a local-realistic
theory are said to be nonlocal. One way of detecting nonlocality of a certain state $\rho$ is to
show that it violates a Bell inequality. The set of nonlocal states
is complementary to the set of local states whose elements
allow probability distributions of the form (\ref{probdist}) with
(\ref{norm}) for any number of apparatuses $n$. It should be noted that proving locality of a state is generally difficult since, in principle, it has to be shown that any Bell inequality is satisfied or an explicit local realistic description has to be found. Unfortunately, also the problem of determining whether a particular Bell inequality is violated or obeyed for a given entangled state is nontrivial.

In detail, to show that a certain state $\rho$ possesses nonlocal
correlations one has to consider quantum mechanical joint
probabilities
\begin{align}
\label{quantumprob}
P^{QM}(A_a=x;B_b=y)=Tr(\ket{x}_{A_a}\bra{x}_{A_a} \otimes
\ket{y}_{B_b}\bra{y}_{B_b} \cdot \rho)
\end{align}
where $\{\ket{x}_{A_a}\}$ and $\{\ket{y}_{B_b}\}$ are orthonormal
eigenvectors of the corresponding observables of Alice
$\{A_1,\ldots,A_n\}$ and Bob $\{B_1,\ldots,B_n\}$. These
probabilities are then inserted into the quantity $I$ corresponding to a particular Bell inequality (\ref{Bellineq}) which can then be
rewritten in the form
\begin{align}
I=Tr(\mathcal{B}_I \rho)
\end{align}
wherein $\mathcal{B}_I$ is called the Bell operator \cite{Verstraete}.

In order to examine if the inequality is preserved or violated
for a given $\rho$ one has find observables that yield the maximum of $I$, i.e.
\begin{align}
\max I=\max_{\left\{\mathcal{B}_{I}\right\}}Tr\left(\mathcal{B}_{I} \rho \right) \ .
\end{align}
This is in general a high-dimensional nonlinear constrained
optimization problem for which analytic solutions are only known for a few special cases \cite{HorodeckiCHSHNSC, BZNSC}. Most often, the quantity $I$ has to be maximized numerically for each given $\rho$.

Numerical tractability and reliability is closely related to the formulation and parameterization of the problem. Our problem can be brought into a computationally beneficial form in the following way: The outcome of $I$ depends on the choice of the observables $\{ A_a \}$ and $\{ B_b \}$, i.e. the orthonormal bases $\{\ket{k}_{A_a}\}$ and $\{\ket{k}_{B_b}\}$. Since all orthonormal bases are local-unitarily related, our problem is equivalent to determining a set of unitary transformations $\{U_{A_a}\}$  and $\{U_{B_b}\}$ that maximizes $I$. It is unknown whether or not there exists a restrictive set of unitaries that in all cases contains the global maximum: Unbiased multiport beam splitters \cite{CGLMPineq,AcinDurt,Kaszlikowski,Durt,TichyBuchleitner} containing only few parameters were shown to be too restrictive in general \cite{Son}. Consequently, we have to take into account all possible unitary transformations, i.e. the unitary group $\mathcal{U}(d)$. Regarding our problem we utilize the recently introduced 'composite parameterization' of $\mathcal{U}(d)$ \cite{composite} allowing an efficient variation. This parameterization is composed of elementary two-dimensional rotations and one-dimensional phase transformations. Explicitly, using one-dimensional projectors
\begin{align}
P_l=\ket{l}\bra{l}
\end{align}
and generalized anti-symmetric $\sigma$-matrices
\begin{align}
\label{sigmamatrices}
\sigma_{m,n}=-i\ket{m} \bra{n} + i \ket{n} \bra{m} \ ,
\end{align}
constructed with orthonormal basis vectors any unitary operator can be written as\footnote{The sequence of the product is $\prod_{i=0}^{N}A_i=A_0 \cdot A_{1} \cdots A_N$}
\begin{align}
\label{Uc}
U_C=\left[\prod_{m=0}^{d-2} \left(\prod_{n=m+1}^{d-1} \mbox{exp} \left( i P_n \lambda_{n,m} \right) \mbox{exp} \left( i \sigma_{m,n} \lambda_{m,n} \right)  \right) \right] \cdot \left[ \prod_{l=0}^{d-1} \mbox{exp}(i P_l \lambda_{l,l})\right] \ .
\end{align}
The contained $d^2$ real parameters $\lambda_{m,n}$ lie within the ranges $\lambda_{m,n} \in \left[0, 2 \pi \right]$ for $m \geq n$ and $\lambda_{m,n} \in \left[0, \frac{\pi}{2} \right]$ for $m < n$. Gathered in a $d \times d$ matrix
\begin{align}
\label{parametermatrix}
\left(
  \begin{array}{ccc}
    \lambda_{0,0} & \cdots & \lambda_{0,d-1} \\
    \vdots & \ddots & \vdots \\
    \lambda_{d-1,0} & \cdots & \lambda_{d-1,d-1} \\
  \end{array}
\right)
\end{align}
this parameters are to be interpreted as follows: a diagonal entry $\lambda_{m,m}$ represents a global phase transformation for the vector $\ket{m}$, while an off-diagonal entry $\lambda_{m,n}$ represents an operation in the two-dimensional subspace spanned by $\ket{m}$ and $\ket{n}$: An upper right entry $\lambda_{m,n}$ is related to a rotation and the corresponding lower left entry $\lambda_{n,m}$ performs a relative phase shift. In our case we can exploit this structure to conveniently eliminate the $d$ physically irrelevant phases that are related to bases in $\mathbb{C}^d$. That is, starting from the basis that was used to define $U_C$ for the observables, one can remove the right part of the parameterization without discarding any solution, i.e. it suffices consider the transformations
\begin{align*}
U_C=\left[\prod_{m=0}^{d-2} \left(\prod_{n=m+1}^{d-1} \mbox{exp} \left( i P_n \lambda_{n,m} \right) \mbox{exp} \left( i \sigma_{m,n} \lambda_{m,n} \right)  \right) \right] \ .
\end{align*}
Optimal values for the parameters can now be determined via search algorithms such as differential evolution
\cite{DE}, simulated annealing \cite{SA} or the Nelder-Mead method \cite{NelderMead}. With respect to this, the composite parameterization has various advantages compared to several other parameterizations. For instance, varying over $\mathcal{U}(d)$ by using arbitrary hermitian operators $H$ and computing $U=\mbox{exp}\left( i H \right)$ as done in \cite{Son} is computationally expensive \cite{MolerCleve}; thus, inappropriate for high-dimensional problems. Computing $U_C$ however, is simple since it is only a product of the matrices $\mbox{exp} \left( i P_n \lambda_{n,m} \right) \mbox{exp} \left( i \sigma_{m,n} \lambda_{m,n} \right)$, which explicitly read
\begin{align*}
\cos (\lambda_{m,n})\ket{n}\bra{n} &+ \sin (\lambda_{m,n})\ket{n}\bra{m}\\
- e^{i \lambda_{n,m}} \sin(\lambda_{m,n})\ket{m}\bra{n}&+e^{i \lambda_{n,m}} \cos (\lambda_{m,n})\ket{m}\bra{m}\\
+ \sum_{k \neq m,n}& \ket{k}\bra{k} \ .
\end{align*}
Also the feature that the product $\prod_{n=m+1}^{d-1} \mbox{exp} \left( i P_n \lambda_{n,m} \right) \mbox{exp} \left( i \sigma_{m,n} \lambda_{m,n} \right)$ leaves the subspace spanned by the basis vectors $\{\ket{0},..,\ket{m-1}\}$ invariant for a fixed $m$ (see \cite{composite}) improves the speed of the computation for large $d$. Efficiently, we
multiply matrices that are smaller than $d \times d$ and in each step $m \rightarrow m+1$ the dimension is increased by one. Another reason to use this particular parameterization besides its computational benefits is the fact that the parameters can directly be related to experimental setups, that is the upper right entries of (\ref{parametermatrix}) correspond to settings of beam splitters and the lower left entries to phase shifters.

\section{The CGLMP--Bell inequality}\label{sectionCGLMP}
A relevant Bell inequality for bipartite systems of dimension $d \times d$ is the Collins--Gisin--Linden--Massar--Popescu--Bell inequality \cite{CGLMPineq}. For $n=2$ observables on each side $A_1, A_2$ and $B_1, B_2$ it was shown that
\begin{align}
\label{CGLMPineq}
&I_d = \sum_{k=0}^{ \left\lfloor  d/2 \right\rfloor -1}\bigl(1-\frac{2k}{d-1}\bigr) \times \\
\bigl\{+\bigl[&P(A_1=B_1+k)+P(B_1=A_2+k+1)\nonumber \\
+&P(A_2=B_2+k)+P(B_2=A_1+k)\bigr] \nonumber \\
-\bigl[&P(A_1=B_1-k-1)+P(B_1=A_2-k) \nonumber\\
+&P(A_2=B_2-k-1)+P(B_2=A_1-k-1)\bigr]\bigr\} \nonumber
\end{align}
with $P(A_a=(B_b+k)\mbox{ mod }d)=\sum_{j=0}^{d-1}{P(A_a=(j+k)\mbox{ mod }d, B_b=j)}$
is bounded by $2$ for local-realistic theories. This inequality can be seen as a generalization of the well known CHSH--Bell inequality \cite{CHSH} since for $d=2$ they are equivalent. The CGLMP-Bell inequality was proven to correspond to facets of local-realistic correlations implied by (\ref{probdist}) and (\ref{norm}) for $n=2$ observables and therefore belongs to the class of tight Bell inequalities \cite{Masanes}. Some known results on the CGLMP--inequality can be used to test the power and reliability of our proposed algorithm. As shown in \cite{CGLMPineq} with a supposed set of optimal observables\footnote{There exists no analytic proof that they are optimal.} one can attain the values
\begin{align}
\label{maxvalue}
I_d&=\frac{2}{d^2}\sum_{k=0}^{[\frac{d}{2}]-1}\left(1-\frac{2k}{d-1}\right)\left(\frac{1}{\sin^2\left(\frac{\pi}{d}(k+\frac{1}{4})\right)}-\frac{1}{\sin^2\left(\frac{-\pi}{d}(k+\frac{3}{4})\right)}\right) \end{align}
for the maximally entangled state $\frac{1}{\sqrt{d}}\sum_{s=0}^{d-1}\ket{s} \otimes \ket{s}$. By maximizing $I_d$ over all observables, i.e. the $4(d^2-d)$ involved parameters $\lambda_{m,n}$ as described in the previous section using the Nelder-Mead method we have been able to confirm these values up to dimension $d=40$ with an accuracy of $10^{-6}$. Thus, the maximum of $I_d$ increases with dimension $d$ and lies within $I_2=2.82843$ for $d=2$ and $I_d=2.96981$ which is reached for large $d$. For all investigated dimensions the algorithm has required only a few runs to find the value given by (\ref{maxvalue}) due to the robustness of the Nelder-Mead method. Similar approaches utilizing other parameterization also confirm (\ref{maxvalue}), however, since they are computationally less tractable they do not reach $d=40$ but only $d=5$ as in \cite{Son,ZohrenGill} or $d=9$ as in \cite{HuDeng}. Note that there are also techniques for deriving upper bounds on $I_d$ \cite{Liang} using semi-definite programming \cite{SDP} which can be used for small $d$ to prove that the found values are indeed global maxima. By maximizing the largest eigenvalue of $\mathcal{B}_{I_d}$ we could also confirm the results in \cite{AcinDurt,ZohrenGill}, namely that for $d\geq3$ the maximal violation of the CGLMP-inequality is attained with non--maximally entangled states.
\section{The state space geometry of entanglement and CHSH--Bell inequality violation for
bipartite qubits}\label{visCHSH}
Now, we come to the main issue of this paper, namely comparing entanglement and quantum nonlocality using geometric representations of the state space.
For a single qubit such a representation is given by the three-dimensional Bloch vector (for generalized Bloch vectors see \cite{BK-Bloch}). In a similar way, any density matrix of a bipartite qubit system can be written as
\begin{align}
\rho=\frac{1}{4}( \mathbbm{1}\otimes\mathbbm{1} + \vec{a}\cdot\vec{\sigma}\otimes\mathbbm{1} + \mathbbm{1}\otimes \vec{b} \cdot \vec{\sigma} + \sum_{m,n=1}^{3}c_{mn}\sigma_m\otimes \sigma_n )
\end{align}
with two three-dimensional real vectors $\vec{a}$ and $\vec{b}$ related to local properties and a real $3 \times 3$ matrix $c_{mn}$ related to correlations. With regard to studying entanglement and nonlocality, we are not interested in the local properties of the system. Hence, local unitary transformations $U_A\otimes U_B$ can be used to diagonalize the matrix $c_{mn}$ (for more details on this and the following steps see \cite{HorodeckiGEOMETRY,BertlmannGEOMETRY}). The diagonal entries can be regarded as components of a three-dimensional real vector $\vec{c}$. Due to the non-negativity condition $\rho \geq 0$ the components of this vector must obey the inequalities\footnote{These are necessary conditions for non-negativity. They are also sufficient when $\vec{a}=0$ and $\vec{b}=0$.}
\begin{align*}
&1+c_{1}+c_{2}-c_{3}\geq0 \ ,\\
&1-c_{1}-c_{2}-c_{3}\geq0 \ ,\\
&1+c_{1}-c_{2}+c_{3}\geq0 \ ,\\
&1-c_{1}+c_{2}+c_{3}\geq0 \ .
\end{align*}
These constraints restrict $\vec{c}$ to lie within a regular tetrahedron spanned by the projectors of the four well-known Bell
states $\ket{\Psi^+},\ket{\Psi^-},\ket{\Phi^+}$ and $\ket{\Phi^-}$. This tetrahedron represents the core of the state space we are interested in and for it, we determine the regions that are entangled or nonlocal, respectively.

For $d=2$ it was proven that the PPT (positive under partial
transposition \cite{PPT}) criterion is necessary and sufficient, i.e. iff all eigenvalues of a partially transposed density matrix are non-negative then the state is separable (for
$d\geq3$ it is only a necessary criterion). By means of this criterion one finds that all states outside the octahedron spanned by the points
$\vec{c}_{1/2}=(0,0,\pm1),\vec{c_{3/4}}=(0,\pm1,0)$ and
$\vec{c_{5/6}}=(0,0,\pm1)$ are entangled. For $\vec{a}=0$ and $\vec{b}=0$ this is necessary and sufficient for entanglement.

\begin{figure}
\centering
\setlength{\unitlength}{0.55bp}
  \begin{picture}(350, 420)(0,0)
  \put(0,0){\includegraphics[scale=0.4]{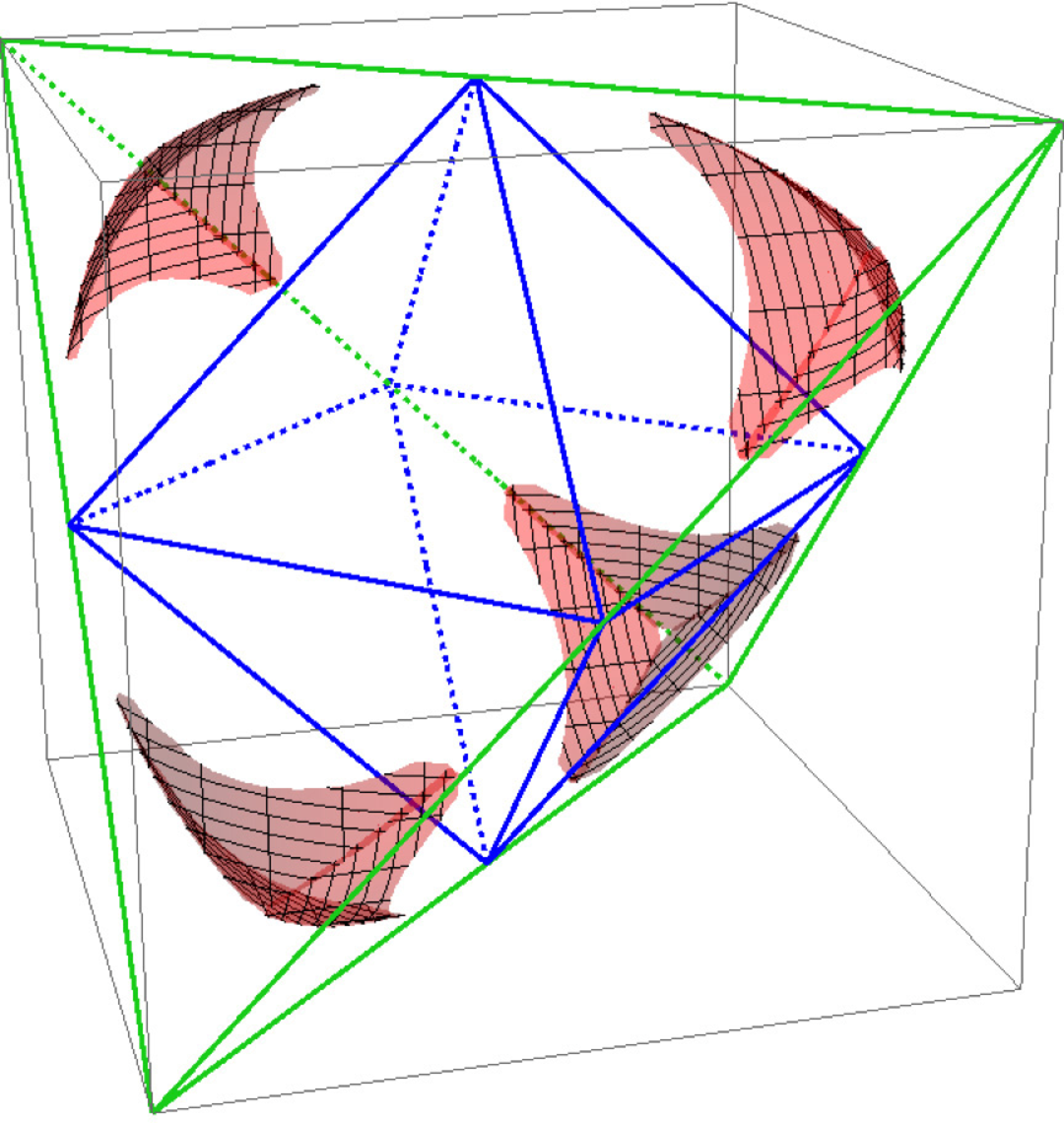}}

  \put(0,20){\fontsize{8}{20}\selectfont \makebox(0,0)[]{$\ket{\Psi^-}\bra{\Psi^-}$\strut}}
  \put(320,170){\fontsize{8}{20}\selectfont \makebox(0,0)[]{$\ket{\Psi^+}\bra{\Psi^+}$\strut}}
  \put(350,400){\fontsize{8}{20}\selectfont \makebox(0,0)[]{$\ket{\Phi^+}\bra{\Phi^+}$\strut}}
  \put(-10,410){\fontsize{8}{20}\selectfont \makebox(0,0)[]{$\ket{\Phi^-}\bra{\Phi^-}$\strut}}
   \end{picture}
  \caption{(Color online) Illustration of the tetrahedron (green) spanned by the four Bell states located in the corners of the cube. By means of the PPT criterion one finds that all states outside this octahedron (blue) are entangled. States beyond the meshed surfaces (red) violate the CHSH--Bell inequality.}
  \label{tetranl}
\end{figure}
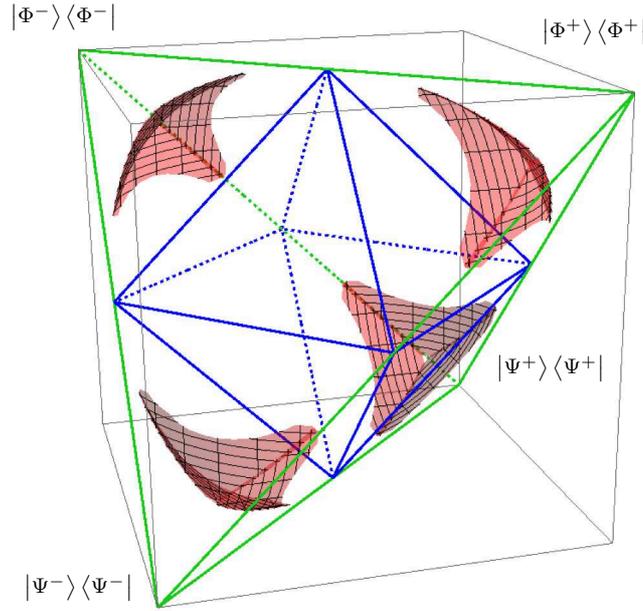

Now, we complete this geometric picture introduced in \cite{HorodeckiGEOMETRY,BertlmannGEOMETRY} by adding the geometry of quantum nonlocality. States of the tetrahedron that violate the CHSH--Bell inequality (CGLMP--inequality for $d=2$) can be determined analytically. According to the criterion given in \cite{HorodeckiCHSHNSC} the maximal attainable value for $I_2$ is $2\sqrt{\lambda_1+\lambda_2}$, where $\lambda_1$ and $\lambda_2$ are the two largest eigenvalues of the matrix $U_{\rho}$ having the components $u_{mn}=\sum_{k=1}^{3} c_{km}c_{kn}$. Consequently, since in our case $c_{mn}$ is a diagonal matrix with the entries $c_1, c_2$ and $c_3$ the Bell inequality is violated iff at least one
of the three inequalities
\begin{align*}
c_{1}^2+c_{2}^2 > &1 \ ,\\
c_{1}^2+c_{3}^2 > &1 \ ,\\
c_{2}^2+c_{3}^2 > &1 \
\end{align*}
holds. This means that each of the inequalities defines a
cylinder with radius $1$ and if a density matrix lies outside of one it is
nonlocal. The union of the exterior regions of these three cylinders
belongs to the set of nonlocal states. The tetrahedron including the boundaries of separability and CHSH violation is visualized in Fig.~\ref{tetranl}. It
descriptively demonstrates that not all entangled states violate the
CHSH inequality. Note that it is also known that all entangled bipartite qubit states can be distilled and therefore do contain so-called
hidden nonlocality \cite{Popescu}.

\section{The state space geometry of bipartite qudits - the magic
simplex $\mathcal{W}$}\label{Wsimplex}
\begin{figure}

\centering
  \setlength{\unitlength}{1.2bp}%
  \begin{picture}(182.42, 154.54)(0,0)
  \put(0,0){\includegraphics[scale=1.2]{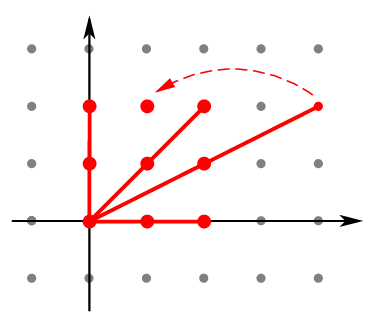}}
  \put(152.18,35.38){\fontsize{12}{14}\selectfont \makebox(0,0)[]{l - position\strut}}
  \put(28.42,115.49){\rotatebox{90.00}{\fontsize{11.38}{13.66}\selectfont \makebox(0,0)[]{k - momentum\strut}}}
  \end{picture}%
  \caption{(Color online) Illustration of the finite discrete classical phase space for $d=3$
  of the simplex $\mathcal{W}$. All possible complete lines
  through the point $(0,0)$ for $d=3$ are drawn, representing one class of states which have the
  same geometry concerning separability and (non-)locality. Lines
  can be completed by points with $k,l \notin \{0,..,d-1\}$ because of the periodicity of the Weyl operators
  implying $P_{k,l}=P_{k+m \cdot d,l+n \cdot d}$ for all $n,m \in \mathbb{Z}$.}
  \label{ps}
\end{figure}

A generalization of the tetrahedron for bipartite qudit systems for arbitrary dimension $d$ was introduced in \cite{BHNsimplexqutrits,BHNsimplex}. Analogously to the qubit case, a regular simplex can be constructed using mutually
orthogonal generalized Bell states. This so-called
``magic simplex'' is given by the set of states
\begin{align}
\mathcal{W}=\{\sum_{k,l=0}^{d-1} c_{k,l}P_{k,l} \mbox{ \ }  |  \mbox{ \ }  c_{k,l}\geq 0,\sum_{k,l=0}^{d-1}c_{k,l}=1\} \ ,
\end{align}
where $P_{k,l}=\ket{\Omega_{k,l}}\bra{\Omega_{k,l}}$ are the
projectors of $d^2$ Bell-type states generated by applying the
Weyl operators \begin{align} W_{kl}=\sum_{s=0}^{d-1}e^{\frac{2\pi i
}{d}sk}\ket{s}\bra{(s+l)\ \mbox{mod}\ d}
\end{align}
with $k,l \in \left\{0,...,d-1\right\}$ on the maximally entangled
state
\begin{align}
\ket{\Omega_{0,0}}=\frac{1}{\sqrt{d}}\sum_{s=0}^{d-1}\ket{s}\otimes\ket{s} \ ,
\end{align}
i.e.
\begin{align}
\ket{\Omega_{k,l}}=\left(W_{k,l}\otimes\mathbbm{1}\right)\ket{\Omega_{0,0}} \ .
\end{align}
The simplex is a convex set located in a $d^2-1$ dimensional
hyperplane in a $d^2$ dimensional real vector space of hermitian
operators spanned by the operators $\{P_{k,l}\}$. Unlike as for two qubits, where every state $\rho$ has a representative in the tetrahedron, for $d\geq3$ not every $\rho$ on $\mathbb{C}^d \otimes \mathbb{C}^d$ can be related to an element of the simplex \cite{BHNsimplexqutrits,BHNsimplex}. However, the derived class of states is of special importance in quantum key distribution \cite{securityQKD} and entanglement distillation protocols \cite{distillqudits}. Moreover, these states are frequently studied in the context of (bound) entanglement \cite{BHNsimplexqutrits2,BKbound,Chruscinski1,Chruscinski2,BKrealignment} and entanglement measures \cite{HHK1,BFreiburg} due their interesting features. A detailed discussion on the properties of $\mathcal{W}$ and on how it is embedded in the state space can be found in
\cite{BHNsimplexqutrits,BHNsimplex}. For the subsequent
investigations let us note that local (anti-)unitary
transformations $U_A \otimes U_B$ mapping $\mathcal{W}$ onto itself are related to symmetries of $\mathcal{W}$, since equivalence classes
\begin{align*}\left[ \rho \right]=\{ \rho' \in \mathcal{W} |
\rho'=U_A\otimes U_B\; \rho\; U_A^{\dagger} \otimes
U_B^{\dagger}\}\end{align*} share the same properties with respect to separability and (non-)locality.
The set of all symmetries can be represented with a discrete classical phase space, see Fig.~\ref{ps}.
From Theorem $9$ of \cite{BHNsimplex} one can infer that states with
components $\{c_{k,l}\}$ and $\{c'_{k',l'}\}$, respectively, belong to the same
equivalence class iff there exists a phase space
transformation of the form
\begin{align}
\label{phasetrans}
\begin{pmatrix} k \\ l \end{pmatrix} = \begin{pmatrix} m & n \\ p&q \end{pmatrix} \begin{pmatrix} k' \\ l' \end{pmatrix} + \begin{pmatrix} j \\ r \end{pmatrix} \hspace{1.5cm} M:=\begin{pmatrix} m & n \\ p&q \end{pmatrix}
\end{align}
with $\det M=1$ or
$\det M =d-1$ such that
$c_{k,l}=c'_{k,l}$ with $m,n,p,q,j,r\in \mathbb{Z}$.

Quantum nonlocality has so far not been investigated for the magic simplex $\mathcal{W}$. In the present paper we aim to determine the states of the magic
simplex $\mathcal{W}$ that violate the CGLMP--Bell inequality by
using the novel approach introduced in Sec.~\ref{nonlocalstates}. In particular, we want to exactly specify the implied boundaries, i.e. states $\rho$ that obey
\begin{align*}\max_{\{\mathcal{B}_{I_d}\}}Tr(\mathcal{B}_{I_d}\;\rho) =
2 \ .
\end{align*}
For arbitrary families of states this is in general difficult since it has to be done iteratively in small regions until a required precision is reached. However, for the considered classes of states one can exploit that the trace of the CGLMP--Bell operator vanishes, i.e. $Tr(\mathcal{B}_{I_d})=0$ which is a consequence of the specific form of the CGLMP--Bell inequality: The quantity $I_d$ (\ref{CGLMPineq}) is composed of the probabilities $P(A_a=(B_b+k)\mbox{ mod }d)=\sum_{j=0}^{d-1}{P(A_a=(j+k)\mbox{ mod }d, B_b=j)}$ and when it is rewritten as $Tr(\mathcal{B}_{I_d} \rho)$ every $P(A_a=x, B_b=y)$ corresponds to a one-dimensional projector $\ket{x}_{A_a}\bra{x}_{A_a} \otimes \ket{y}_{B_b}\bra{y}_{B_b}$ which has trace $1$. Since there are equally many projectors with positive and negative prefactors in every term of the sum in (\ref{CGLMPineq}) and due to the linearity of the trace it follows that $Tr(\mathcal{B}_{I_d})=0$.

Consequently, for states of the form $\rho=\frac{1-\nu}{d^2}\mathbbm{1}_{d^2}+\nu \tau$, i.e. a particular state $\tau$ mixed with uncolored noise $\frac{1}{d^2}\mathbbm{1}_{d^2}$ we can exploit that
\begin{align}
\max_{\{\mathcal{B}_{I_d}\}}Tr(\mathcal{B}_{I_d}\rho)&=
\max_{\{\mathcal{B}_{I_d}\}}Tr(\mathcal{B}_{I_d}\left[\frac{1-\nu}{d^2}\mathbbm{1}_{d^2}+\nu\;\tau\right]) \nonumber \\
&= \max_{\{\mathcal{B}_{I_d}\}}Tr(\mathcal{B}_{I_d}\nu\; \tau) \nonumber \\
&= \nu \max_{\{\mathcal{B}_{I_d}\}}Tr(\mathcal{B}_{I_d}\; \tau) \label{addnoise} \ ,
\end{align}
which implies that if the
maximal value of $\max I_d(\tau)=\max_{\mathcal{B}_{I_d}}Tr(\mathcal{B}_{I_d}\tau)$ is
known then the parameter value $\nu=2/\max I_d(\tau)$ for $\rho$ yields a state on the boundary of CGLMP--Bell inequality violation.

In order to compare our new results on the geometry of quantum nonlocality with the geometry of entanglement we use several results and techniques of precedent publications on the magic simplex $\mathcal{W}$. For a detailed discussion of how to decide separability for the simplex states via the PPT criterion, matrix realignment, optimal
entanglement witnesses and entanglement measures we refer the reader to \cite{BHNsimplexqutrits,BHNsimplex,BHNsimplexqutrits2,BKbound,BKrealignment,BFreiburg}.
\section{The state space geometry of entanglement and CGLMP--Bell inequality violation for
bipartite qutrits}\label{visqutrits}
In the following we investigate and
illustrate low-dimensional sections of the $\mathcal{W}$-simplex for
bipartite qutrits ($d=3$). As a first example
consider the so-called isotropic states
\begin{align} \label{isostate} \rho=\frac{1-\alpha}{9} \mathbbm{1}_9 + \alpha P_{k,l}\;.\end{align}
The phase space transformation rules (\ref{phasetrans}) imply that
all such one parameter states with arbitrary $k,l\in\{0,..,2\}$ but
same $\alpha$ have the same properties in terms of separability/entanglement and (non-)locality. Any state of this form is separable for
$\frac{1}{4} \geq   \alpha \geq 0$ and entangled for $\alpha >
\frac{1}{4}$ (see \cite{Narnhofer} and references therein). Using (\ref{addnoise}) and the compliance of our numerical results with (\ref{maxvalue}) we find that the CGLMP--Bell
inequality is violated for $\alpha>\frac{1}{2}(6\sqrt{3}-9)$.

\begin{figure}
\centering
\includegraphics[scale=0.35]{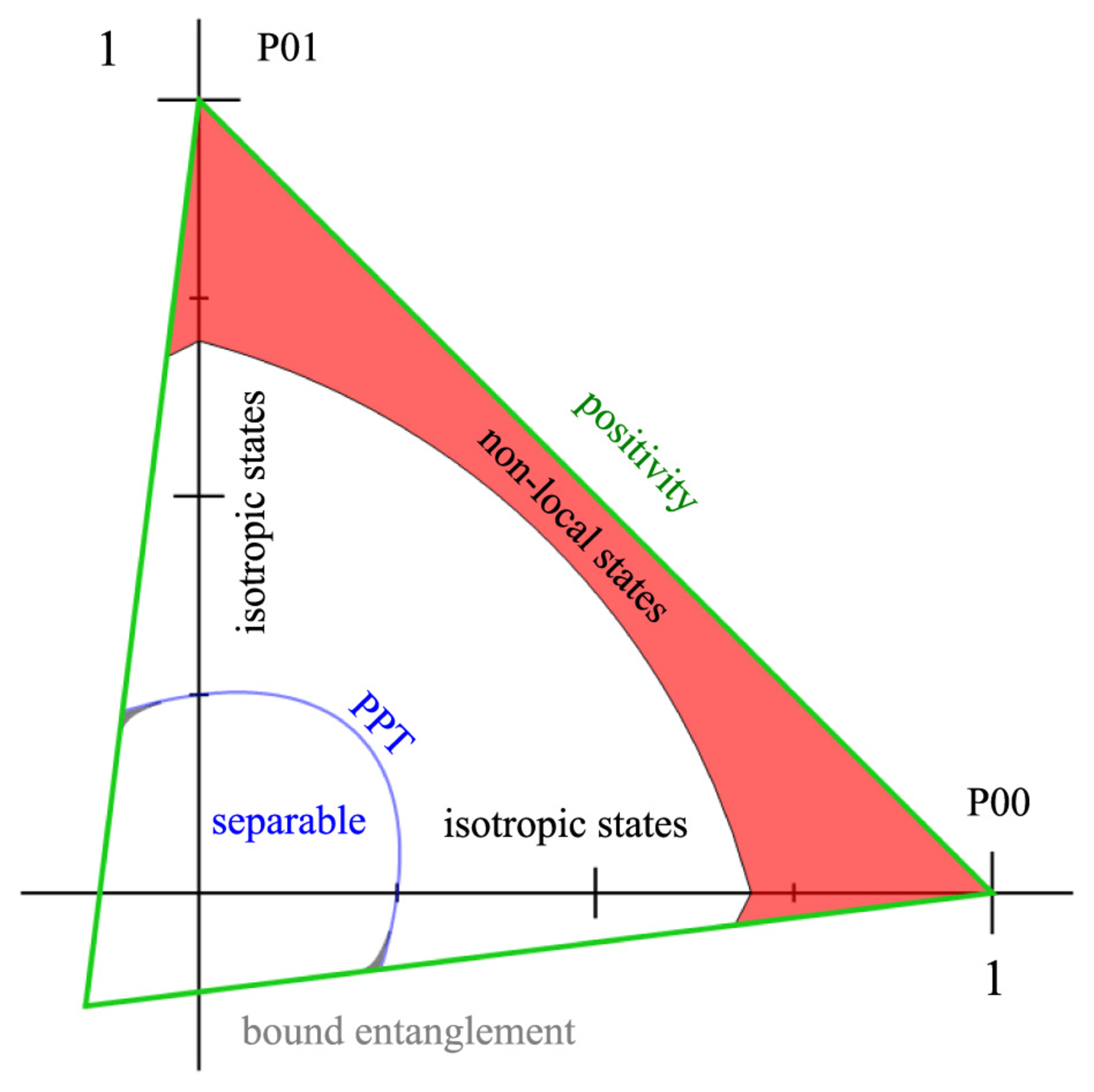}
\setlength{\unitlength}{0.63bp}%
  \begin{picture}(0, 0)(0,0)
\put(-10,50){\fontsize{12}{14}\selectfont \makebox(0,0)[]{$\alpha$ \strut}}
\put(-280,340){\fontsize{12}{14}\selectfont \makebox(0,0)[]{$\beta$ \strut}}
\end{picture}
\caption{(Color online) Illustration of the state
$\rho=\frac{1-\alpha-\beta}{9} \mathbbm{1}_9 + \alpha P_{0,0} +
\beta P_{0,1}$. All states lie within the (green) triangle which
corresponds to the border of positivity. The (blue) ellipse
corresponds to the PPT border, i.e. all PPT states lie inside.
As shown in \cite{BHNsimplexqutrits} if one parameter is
negative, there is also a small (gray filled) region of bound entanglement. States $\rho$ in the (red) filled area violate the
CGLMP--Bell inequality $(\max I_3(\rho)>2)$. Interestingly, the
geometry given by the CGLMP--Bell operator changes its shape (from a circle to a line) at the transition from positive to negative parameters.}
  \label{NL1}
\end{figure}
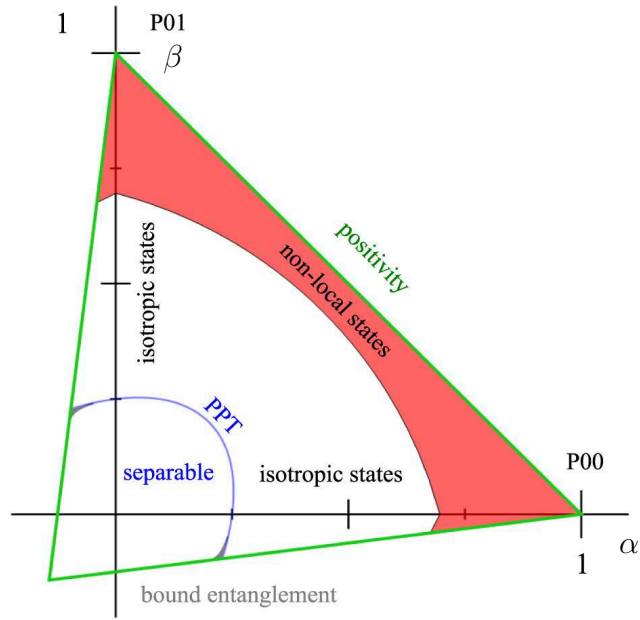

\begin{figure}
\centering
\includegraphics[scale=0.42]{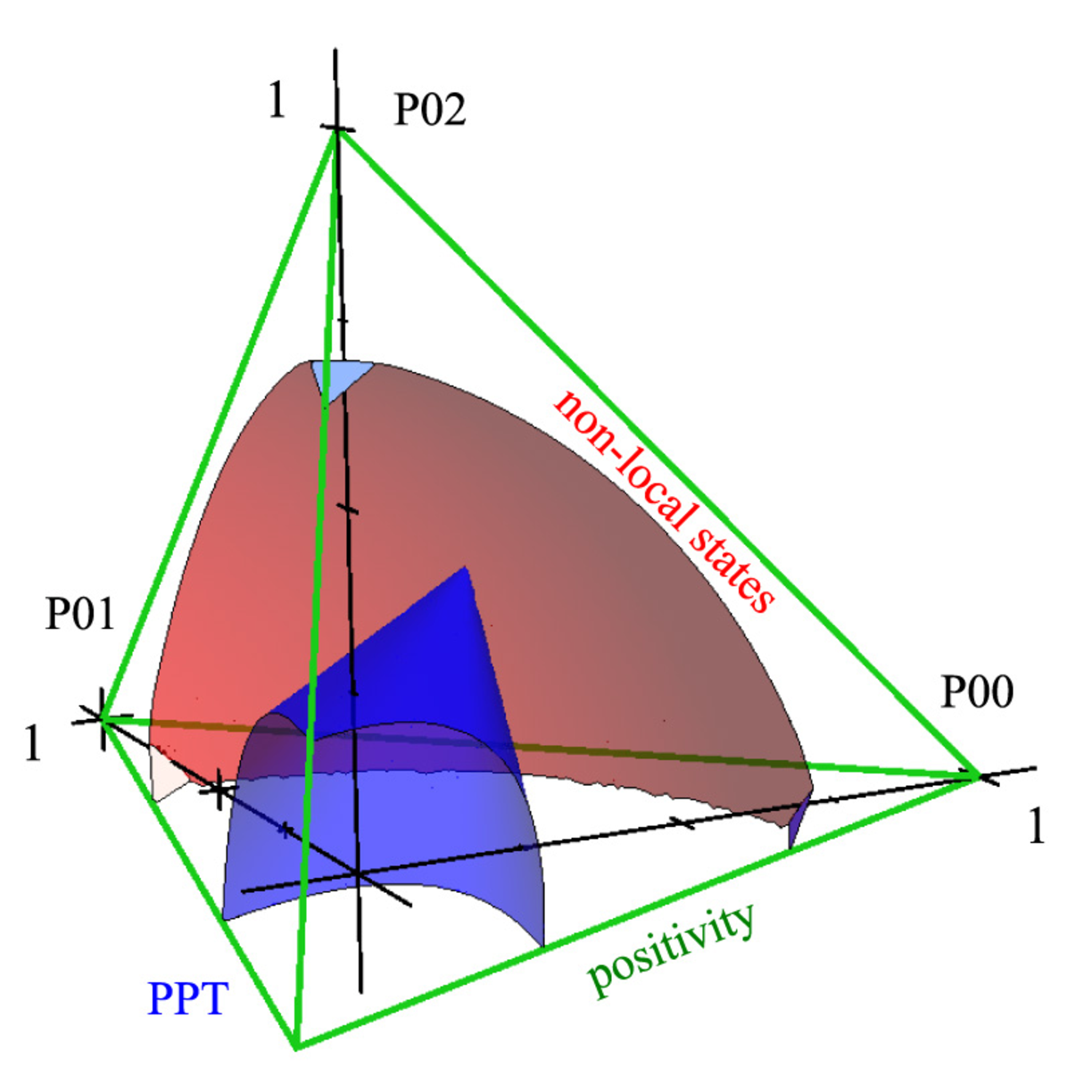}
\setlength{\unitlength}{1cm}
  \begin{picture}(0, 0)(0,0)
\put(-8.8,3.3){\fontsize{12}{14}\selectfont \makebox(0,0)[]{$\beta$ \strut}}
\put(-5.8,8.3){\fontsize{12}{14}\selectfont \makebox(0,0)[]{$\gamma$ \strut}}
\put(-0.2,3.0){\fontsize{12}{14}\selectfont \makebox(0,0)[]{$\alpha$ \strut}}
\end{picture}
  \caption{ (Color online)
Illustration of a three dimensional state subspace given by bipartite
qutrit line states $\rho_{\textrm{line}}(\alpha,\beta,\gamma)$,
Eq.~(\ref{line}). As for bipartite qubit states, Fig.~\ref{tetranl},
all states have to be within a (green) tetrahedron (positivity
condition). The boundary of PPT states forms a (blue) cone and
thus all states beyond this surface are entangled. The tip of the
cone touches the surface of positivity at
$\alpha=\beta=\gamma=\frac{1}{3}$. States $\rho$ beyond the shaded
(light red/blue) surface violate the CGLMP--Bell inequality $(\max
I_3(\rho)>2)$. Again, the geometric form changes from a sphere to a plane when one or more parameters become negative.} \label{lineNL}
\end{figure}
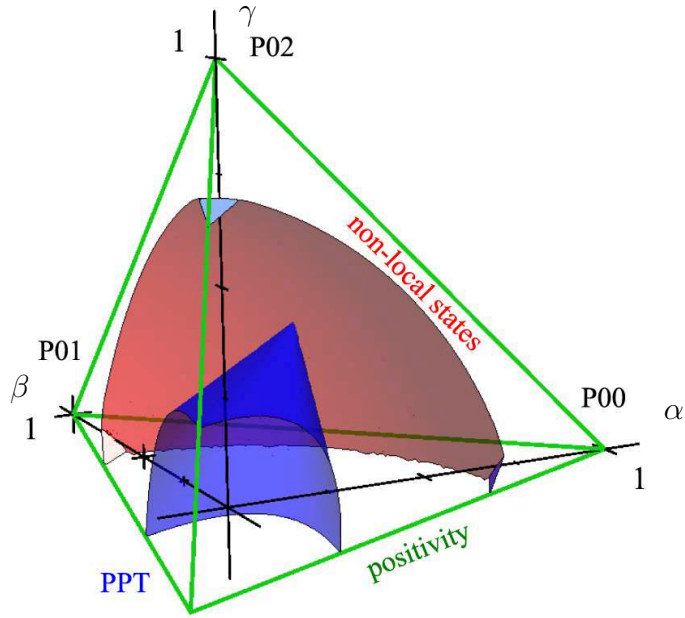

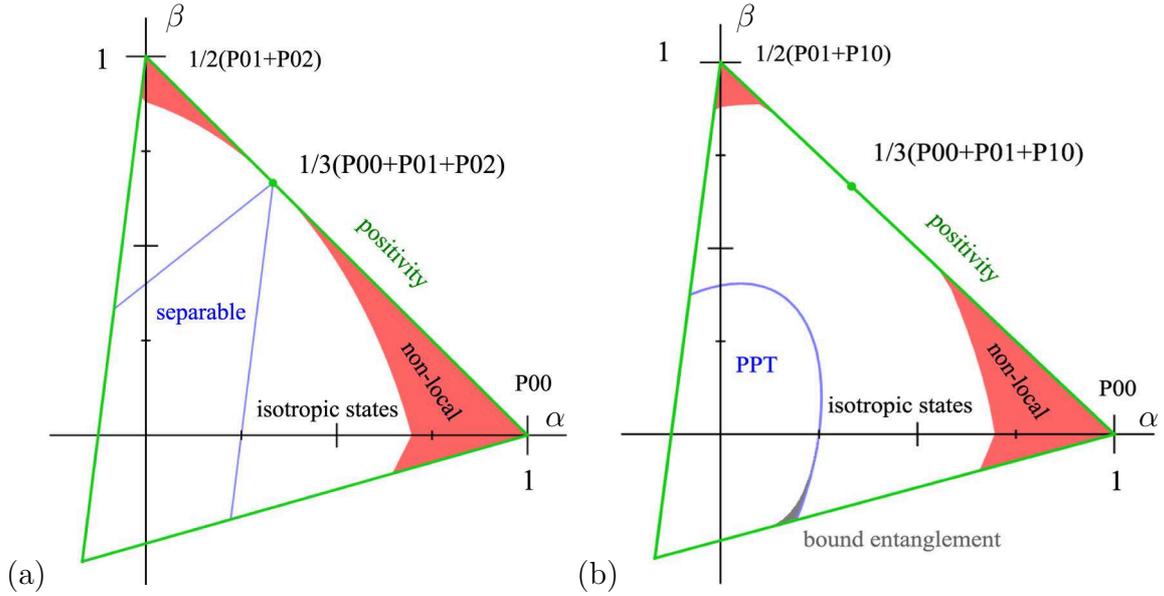
\begin{figure*}
\centering (a)\includegraphics[scale=0.40]{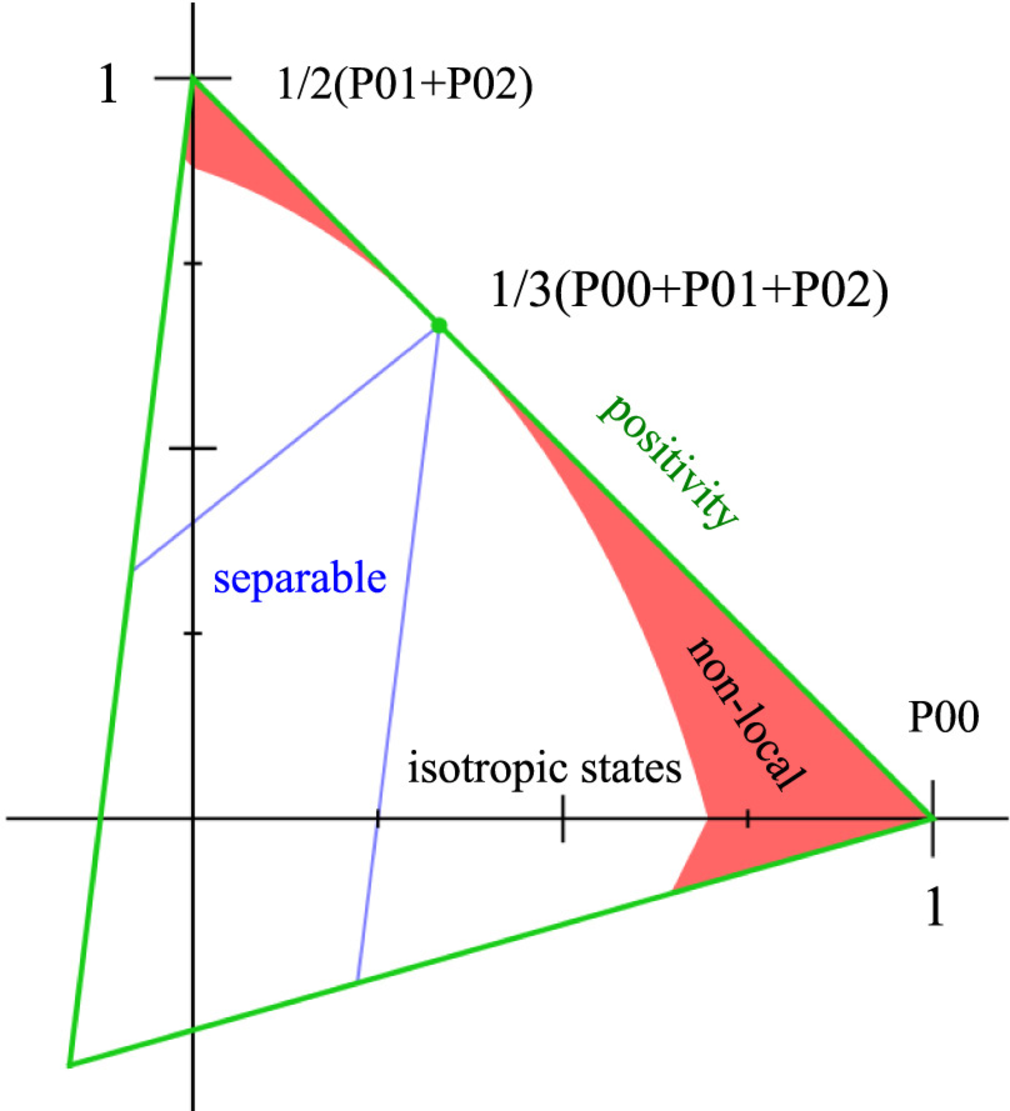}
(b)\includegraphics[scale=0.40]{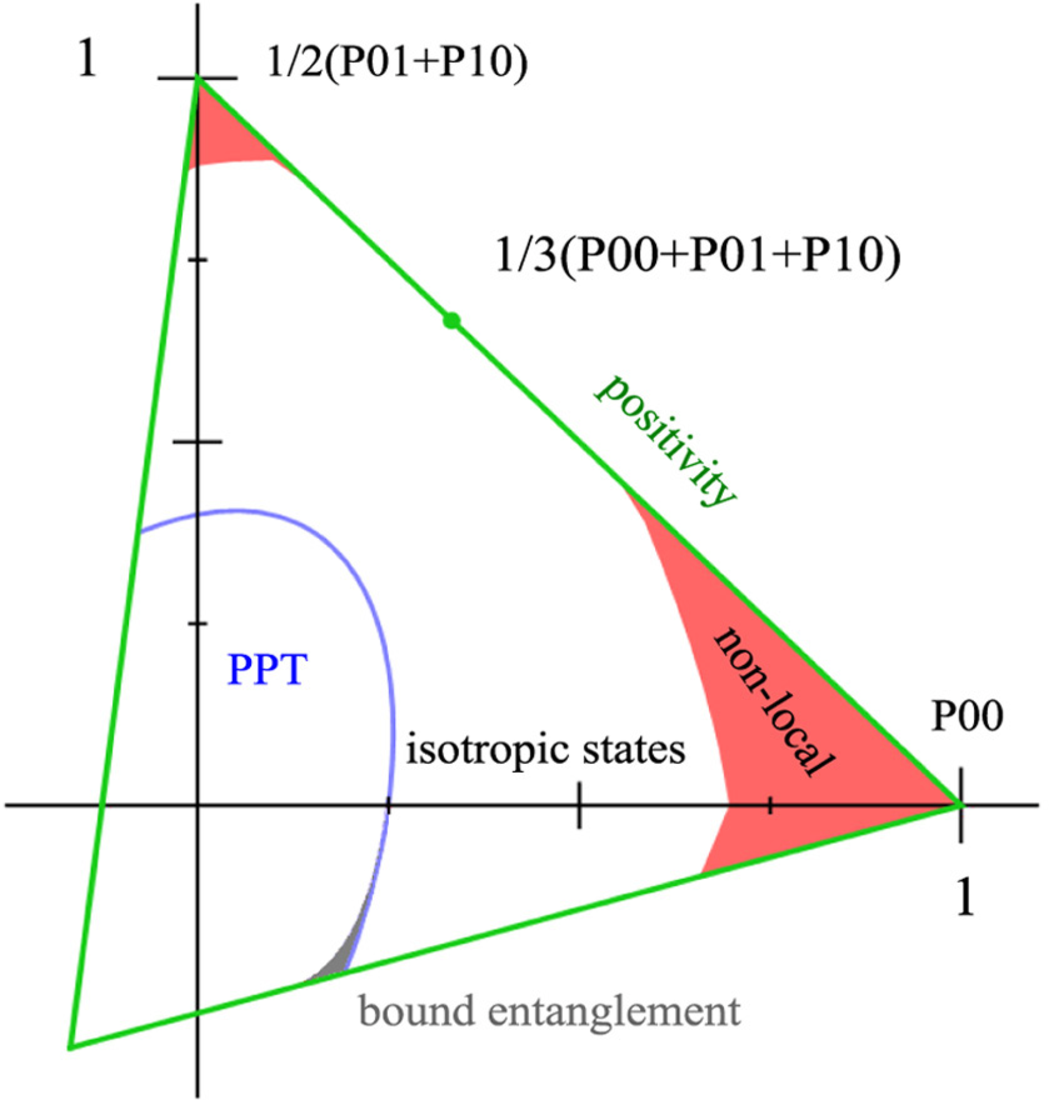}
\setlength{\unitlength}{1cm}%
  \begin{picture}(0, 0)(0,0)
\put(-8.0,2.2){\fontsize{12}{14}\selectfont \makebox(0,0)[]{$\alpha$ \strut}}
\put(-0.2,2.2){\fontsize{12}{14}\selectfont \makebox(0,0)[]{$\alpha$ \strut}}
\put(-13.0,7.5){\fontsize{12}{14}\selectfont \makebox(0,0)[]{$\beta$ \strut}}
\put(-5.5,7.5){\fontsize{12}{14}\selectfont \makebox(0,0)[]{$\beta$ \strut}}
\end{picture}
  \caption{(Color online) Comparison of the class of line states
  $\rho_{\textrm{line}}(\alpha,\frac{\beta}{2},\frac{\beta}{2})$,
  Eq.~(\ref{line}), with the class of off-line states $\rho_{\textrm{off
  line}}(\alpha,\frac{\beta}{2},\frac{\beta}{2})$, Eq.~(\ref{offline}). In the first case, one finds that there is no bound entanglement and that the separable states form a polygon. In the second case, the boundary given by the PPT criterion is curved (blue) and bound entanglement is found for
  $\beta<0$ (grey filled region). The region of nonlocal states (filled red region) is smaller in case the third Bell state does not belong
  to the line formed by the other two Bell states of the mixture.}
  \label{P00P01P10}
\end{figure*}

Next, consider the two-parameter families of states of the form
\begin{align}\rho=\frac{1-\alpha-\beta}{9} \mathbbm{1}_9 + \alpha P_{k,l} +
\beta P_{m,n}\end{align} with arbitrary $k,l,m,n\in\{0,..,2\}$.
Any such state is local-unitarily equivalent to the state
$\rho=\frac{1-\alpha-\beta}{9} \mathbbm{1}_9 + \alpha P_{0,0} +
\beta P_{0,1}$. For this set of states the PPT boundary reads
\begin{align*}8 \alpha^2 + 8 \beta^2 - 11 \beta \alpha + 2 \alpha + 2 \beta -
1 = 0\; ,\end{align*} which was derived by setting the eigenvalues of the partially transposed matrix $\rho^{T_B}$ to zero. There are also bound entangled states in
this set. They can be found through optimal entanglement witnesses
yielding the boundaries \cite{BFreiburg} \begin{align*}&4
\alpha^2-5 \alpha+40
\beta^2+(17 \alpha-14) \beta+1=0\qquad\textrm{and}\\
&4 \beta^2-5 \beta+40 \alpha^2+(17 \beta-14)
\alpha+1=0\;.\end{align*} The boundary of the CGLMP--Bell
violation was obtained by computing $\max I_3(\rho)$ for $200$
equally spaced points on the boundaries of positivity ($\alpha =
\frac{\beta-1}{8}$, $\beta = \frac{\alpha-1}{8}$ and $\beta =
1-\alpha$). Again, (\ref{addnoise}) was exploited to determine the values of $\alpha$ and $\beta$
corresponding to states on the boundary, i.e. $\max_{\mathcal{B}_{I_d}}Tr \left( \mathcal{B}_{I_d}\rho \right)=2$. The result is graphically illustrated in
Fig.~\ref{NL1}. The illustration suggests that the boundary of CGLMP--Bell inequality violation
describes a circle for $\alpha,\beta>0$ and a line if one of the
parameters is negative, i.e. $\alpha<0$ or $\beta<0$. Note that this abrupt change in the shape only appears for the
boundary of CGLMP--Bell inequality violation but not for the boundary of separability.
Conjectures on the exact specifications of these boundaries are
contained in the considerations of the three-parameter
families of states
\begin{align}\rho=\frac{1-\alpha-\beta-\gamma}{9} \mathbbm{1} +
\alpha P_{k,l} + \beta P_{m,n} + \gamma P_{p,q}\end{align} as
the special case $\gamma=0$. Theorem $3$ of \cite{BHNsimplexqutrits} implies that, depending on whether the index pairs
$\{(k,l),(m,n),(p,q)\}$ form a line or not (in the sense of the discrete classical phase space, Fig. 2), this three-parameter
family of states is either local-unitarily equivalent to the state
\begin{align}\label{line}\rho_{\textrm{line}}(\alpha,\beta,\gamma)=\frac{1-\alpha-\beta-\gamma}{9} \mathbbm{1}_9 + \alpha P_{0,0} + \beta P_{0,1} + \gamma P_{0,2}\end{align} or
\begin{align}\label{offline}\rho_{\textrm{off-line}}(\alpha,\beta,\gamma)=
\frac{1-\alpha-\beta-\gamma}{9} \mathbbm{1}_9 + \alpha
P_{0,0} + \beta P_{0,1} + \gamma P_{1,0}\;.
\end{align}
The class of line states are graphically illustrated in
Fig.~\ref{lineNL}, and a particular slice of the off-line states is
visualized in Fig.~\ref{P00P01P10}~(b).

For the states $\rho_{\textrm{line}}(\alpha,\beta,\gamma)$ the PPT boundary is \begin{align*}8 \alpha^2 +
8 \beta^2 + 8 \gamma^2 + 2 \alpha  + 2 \beta + 2 \gamma\nonumber - 11 \alpha \beta  - 11 \alpha \gamma - 11 \beta \gamma  -1=0.\end{align*}
As stated in \cite{BFreiburg} further (bound) entangled states can be revealed using optimal entanglement witnesses which for the boundary of separability written in implicit form yield
\begin{align*}40
\alpha^2+ \alpha (17
\beta+17 \gamma-14) +4 \beta^2+\gamma (4 \gamma-5)-\beta (19 \gamma+5)+1=0 \ , \end{align*} and the equations one gets via permutations with respect to $\alpha,\beta,\gamma$.
The boundary of CGLMP--Bell violation is deduced via (\ref{addnoise}) from the values
$\max I_3(\rho)$ of $1000$ equally spaced states on the boundaries of
positivity ($\alpha = \frac{\beta+\gamma-1}{8}$ (and all parameter
permutations of this term) and $\gamma = 1-\alpha-\beta$). The
resulting points apparently describe a spherical surface which meets
the boundaries of positivity and PPT at the point
$\alpha=\beta=\gamma=\frac{1}{3}$. This sphere appears to be intersected by planes in
the vicinity of the isotropic states, Fig.~\ref{lineNL}.

When taking into account symmetries and the fact that (\ref{maxvalue}) yields the
boundary parameter $\frac{1}{2} (6 \sqrt{3}-9)$ for the isotropic
states one can suppose a sphere with radius $r=\frac{1}{156} (413
\sqrt{3}-558)$ and center at $\alpha=\beta=\gamma=\frac{1}{156} (-361+186 \sqrt{3})$. These
specifications coincide with the numerical data up to the order
$10^{-6}$.

For $\rho_{\textrm{line}}(\alpha,\beta,\gamma)$ the measurement settings given in \cite{CGLMPineq} combined with local
unitary transformations constituted by the Weyl operators yield the illustrated intersecting planes given by the possible parameter permutations of the function
\begin{align} \label{plane} \gamma=\frac{1}{2}(\alpha+\beta+6 \sqrt{3}-9) . \end{align}
To see this one can use the local-unitarily equivalent state $\rho=\frac{1-\alpha-\beta-\gamma}{9} \mathbbm{1}_9 + \alpha P_{0,0} + \beta P_{1,0} + \gamma P_{2,0}$ and write $Tr(\mathcal{B}_{I_3}\rho)=\alpha Tr(\mathcal{B}_{I_3}P_{0,0})+ \beta Tr(\mathcal{B}_{I_3}P_{1,0})+ \gamma Tr(\mathcal{B}_{I_3}P_{2,0})$.\footnote{Note that it is not an error that the values for $P_{1,0}$ and $P_{2,0}$ differ from that of $P_{0,0}$. Different states, even though they are local-unitarily equivalent require different measurement settings.} Here, the measurement bases in Ref.~\cite{CGLMPineq} yield $Tr(\mathcal{B}_{I_3}P_{0,0})=\frac{4}{6\sqrt{3}-9}$ and $Tr(\mathcal{B}_{I_3}P_{1,0})=Tr(\mathcal{B}_{I_3}P_{2,0})=\frac{-2}{6\sqrt{3}-9}$. Thus, by setting $Tr(\mathcal{B}_{I_3}\rho)=2$ we obtain (\ref{plane}) up to a parameter permutation, which in practice can be realized using a local-unitarily modified Bell operator $W_{k,l}^{\dagger}\otimes\mathbbm{1}_3 \ \mathcal{B}_{I_3} \ W_{k,l}\otimes\mathbbm{1}_3$ since $Tr(\mathcal{B}_{I_3}P_{k,l})=Tr(\mathcal{B}_{I_3} \ W_{k,l}\otimes\mathbbm{1}_3 \ P_{0,0} \ W_{k,l}^{\dagger}\otimes\mathbbm{1}_3)=Tr((W_{k,l}^{\dagger}\otimes\mathbbm{1}_3 \ \mathcal{B}_{I_3} \ W_{k,l}\otimes\mathbbm{1}_3) \ P_{0,0})$. For parameter regions in the vicinity of the isotropic states where one parameter is positive and two parameters are equal or less than zero the approach of Sec. \ref{nonlocalstates} does not lead to a better result than that.

For the remaining family of off-line states,
Eq.~(\ref{offline}), the geometric form of the boundary of
CGLMP--Bell violation has a more complex shape. An uncovered
suggestion on the exact form has therefore not been made.
Fig.~\ref{P00P01P10}~(b) gives an impression of the fact that the
boundaries of PPT, as well
as CGLMP--Bell violation have a different shape for the different
types of states. For instance, in contrast to the class of line states where there is only one state
$\alpha=\beta=\gamma=\frac{1}{3}$ on the boundary of positivity
$1-\alpha-\beta-\gamma=0$ that does not violate the CGLMP--Bell
inequality, for the second type $\rho_{\textrm{off-line}}$ there is a whole region of states
for $1-\alpha-\beta-\gamma=0$ that obeys the CGLMP--Bell inequality.
Moreover, in comparison the entire region of nonlocal states in Fig.~\ref{P00P01P10}~(b) is
smaller than in Fig.~\ref{P00P01P10}~(a). For entanglement, one finds the opposite, which is
a intriguing result. It is rather counterintuitive because the
region of separable states of the subset
\begin{align}\label{eqmixed}\rho_{\textrm{line}}(\alpha,\frac{\beta}{2},\frac{\beta}{2})
\end{align} already
is larger than the PPT region of the same subset of off-line states
\begin{align}\rho_{\textrm{off-line}}(\alpha,\frac{\beta}{2},\frac{\beta}{2})
\end{align} which in addition
also contains bound entanglement for $\beta < 0$.

\section{Entanglement measures vs. Bell inequality violations}\label{measure}
In \cite{HHK1,HHmeasure}, an entanglement measure for multipartite qudit systems was introduced, as well as a method how optimal bounds can be derived for it. Here we apply this measure
to bipartite qutrits.

For any bipartite qutrit state $\rho=\sum_i p_i |\psi_i\rangle\langle\psi_i|$, entanglement can be quantified via
\begin{eqnarray}
\mathcal{E}(\rho)&:=&\inf_{\{p_i,\ket{\psi_i}\}} \sum_i p_i\; \left\lbrace
S(\text{Tr}_{A}|\psi_i\rangle\langle\psi_i|)
+S(\text{Tr}_{B}|\psi_i\rangle\langle\psi_i|)\right\rbrace\\&
=&2 \inf_{\{p_i,\ket{\psi_i}\}} \sum_i p_i\;
S(\text{Tr}_{A}|\psi_i\rangle\langle\psi_i|)\;,
\label{measuredef}
\end{eqnarray}
where $S$ is any Renyi entropy. In our case we use the linear entropy $S_L(\rho)=\frac{3}{2}(1-Tr\rho^2)$. In general the infimum,
i.e. the optimal decomposition is not known. The derivation of lower bounds on the measure are based on the observation that
the linear entropy can be rewritten as \cite{HHK1,composite}
\begin{align}
\frac{4}{3}S_L(\rho_A)&=\sum_{k_A<l_A}\sum_{k_B<l_B} \text{Tr}(\ket{\psi}\bra{\psi}\sigma_{k_A,l_A}\otimes\sigma_{k_B,l_B}(\ket{\psi}\bra{\psi})^*\sigma_{k_A,l_A}\otimes\sigma_{k_B,l_B})\\ &=:C_m^2(|\psi\rangle\langle\psi|)
\end{align}
using the anti-symmetric $\sigma$-matrices (\ref{sigmamatrices}). $C_m^2(|\psi\rangle\langle\psi|)$ is called the $m$--concurrence and is an entanglement measure for pure states in its own right. The m-concurrence is generalized to mixed states via the convex roof construction (\ref{measuredef}) to
\begin{align}
\label{mcon1}
C_m^2(\rho):&=\inf_{\{p_i,|\psi_i\rangle\}}\sum_i p_i C^2_m(|\psi_i\rangle\langle\psi_i|)\\
&= \inf_{\{p_i,|\psi_i\rangle\}}\sum_i  p_i \sum_{k_A<l_A}\sum_{k_B<l_B}  \text{Tr}(\ket{\psi_i}\bra{\psi_i}\sigma_{k_A,l_A}\otimes\sigma_{k_B,l_B}(\ket{\psi_i}\bra{\psi_i})^*\sigma_{k_A,l_A}\otimes\sigma_{k_B,l_B}) \,. \nonumber
\end{align}
This expression has the lower bound
\begin{align}
C_m^2(\rho)\geq\sum_{k_A<l_A}\sum_{k_B<l_B}\inf_{\{p_i,|\psi_i\rangle\}}\sum_i p_i \text{Tr}(\ket{\psi_i}\bra{\psi_i}\sigma_{k_A,l_A}\otimes\sigma_{k_B,l_B}(\ket{\psi_i}\bra{\psi_i})^*\sigma_{k_A,l_A}\otimes\sigma_{k_B,l_B}) \ , \nonumber
\end{align}
for which the contained individual infima are known as (see \cite{mintert05,wootters98})
\begin{align}
\inf_{\{p_i,|\psi_i\rangle\}}\sum_i p_i \text{Tr}(\ket{\psi_i}\bra{\psi_i}\sigma_{k_A,l_A}\otimes\sigma_{k_B,l_B}(\ket{\psi_i}\bra{\psi_i})^*\sigma_{k_A,l_A}\otimes\sigma_{k_B,l_B}) =X^2_{k_A,l_A,k_B,l_B}\, ,
\end{align}
with
\begin{align}
\label{bounds}
X_{k_A,l_A,k_B,l_B}:=\max[2\max[\{x^i_{k_A,l_A,k_B,l_B}\}]-\sum_ix^i_{k_A,l_A,k_B,l_B},0]
\end{align}
where $\{x^i_{k_A,l_A,k_B,l_B}\}$ are the square roots of the eigenvalues of
\begin{align}
\rho \ \sigma_{k_A,l_A}\otimes\sigma_{k_B,l_B} \ \rho^* \ \sigma_{k_A,l_A}\otimes\sigma_{k_B,l_B} \ .
\end{align}
The derived bound in general depends on in which basis the state $\rho$ is written and is thus not invariant under local unitaries \cite{composite}. Consequently, the optimal lower bound for a state $\rho$ is given by the maximum over all $\rho'=U_A \otimes U_B \ \rho \ U_A^{\dagger} \otimes U_B^{\dagger}$. This problem can be reduced to a $12$-parameter optimization problem analogously to Sec.~$\ref{nonlocalstates}$. Note that there are also algorithms to numerically derive tight upper bounds on the convex roof \cite{Rothlisberger}, thus we know for which classes of states the exact value of the measure is obtained.
\begin{figure}[htb]
\centering
\includegraphics[scale=0.45]{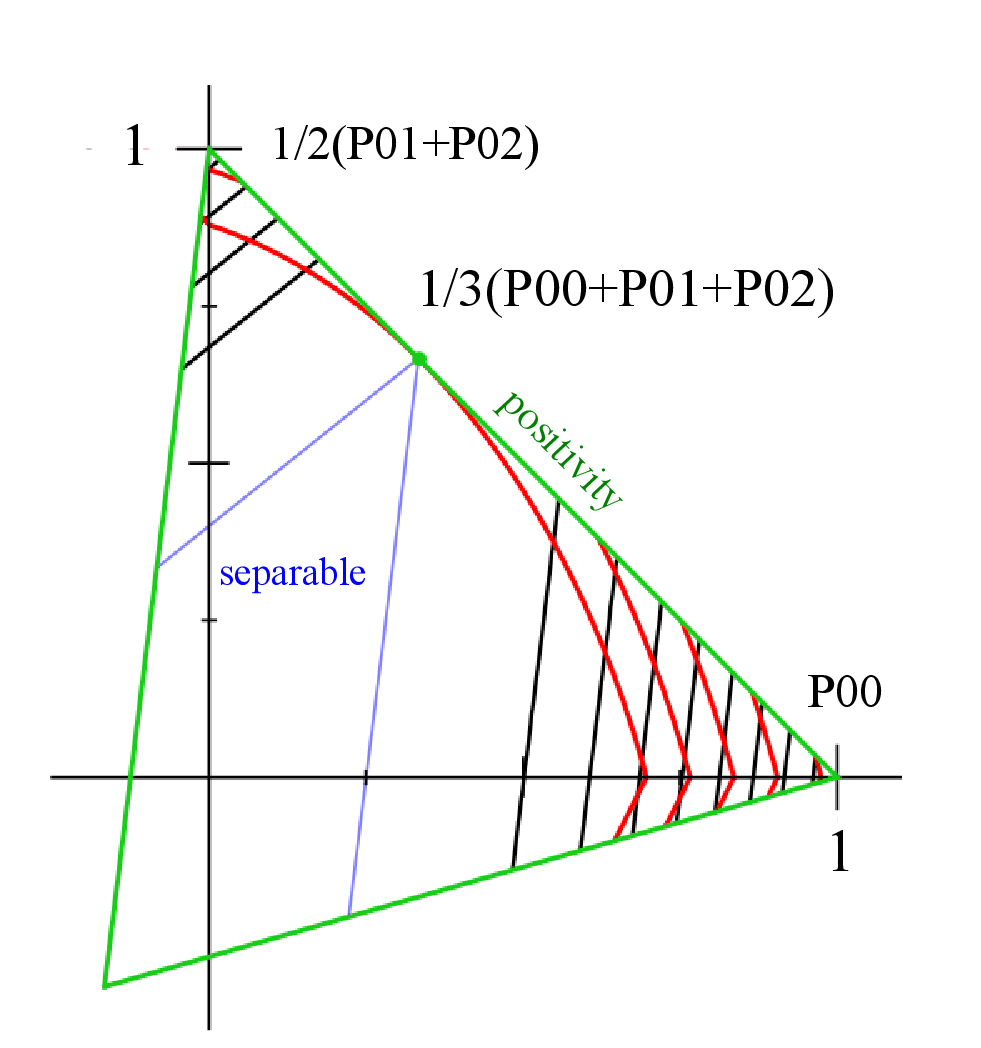}
\setlength{\unitlength}{1cm}%
\begin{picture}(0, 0)(0,0)
\put(-0.8,2.3){\fontsize{12}{14}\selectfont \makebox(0,0)[]{$\alpha$ \strut}}
\put(-5.6,7.3){\fontsize{12}{14}\selectfont \makebox(0,0)[]{$\beta$ \strut}}
\end{picture}
  \caption{(Color online) Comparison of the contour plots of the $m$-concurrence $C_m^2(\rho)$ with Bell inequality violation given by $\max I_{3}(\rho)$ for the  line states $\rho_{\textrm{line}}(\alpha,\frac{\beta}{2},\frac{\beta}{2})$. The black lines parallel to the boundaries of separability correspond to the amount of entanglement, while the red curves correspond to the degree of Bell inequality violation. The fact that both quantities cannot be monotonically related can easily be seen by walking along one of red lines on the right side, i.e. $\alpha  \geq\frac{1}{4}+\frac{\beta}{8}$. For instance, starting from one of the points where a red curve intersects the upper green boundary of positivity, the amount of entanglement increases first until $\beta=0$ but when $\beta$ becomes negative it decreases again.}
  \label{measurecplot}
\end{figure}
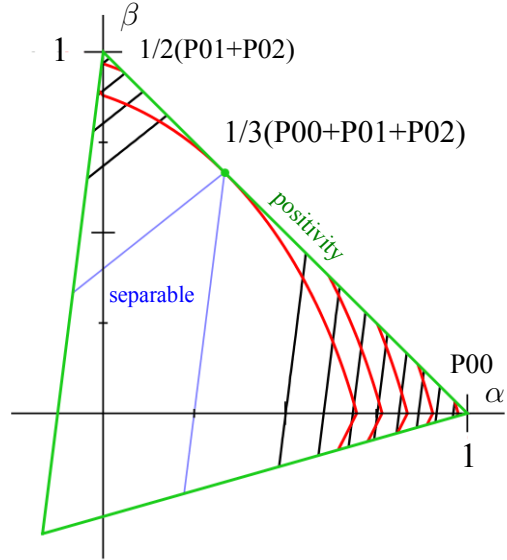

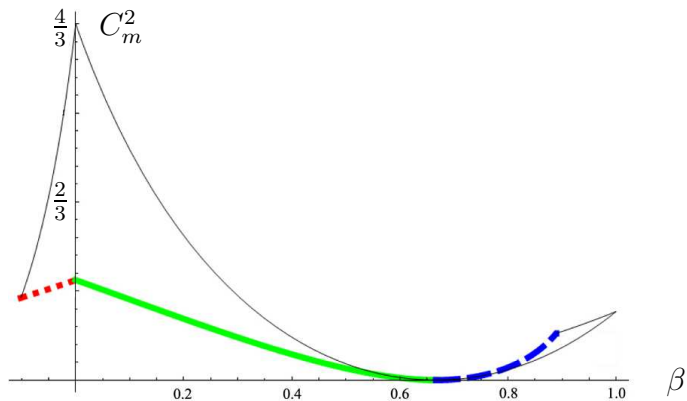
\begin{figure}[htb]
\centering
\includegraphics[scale=0.35]{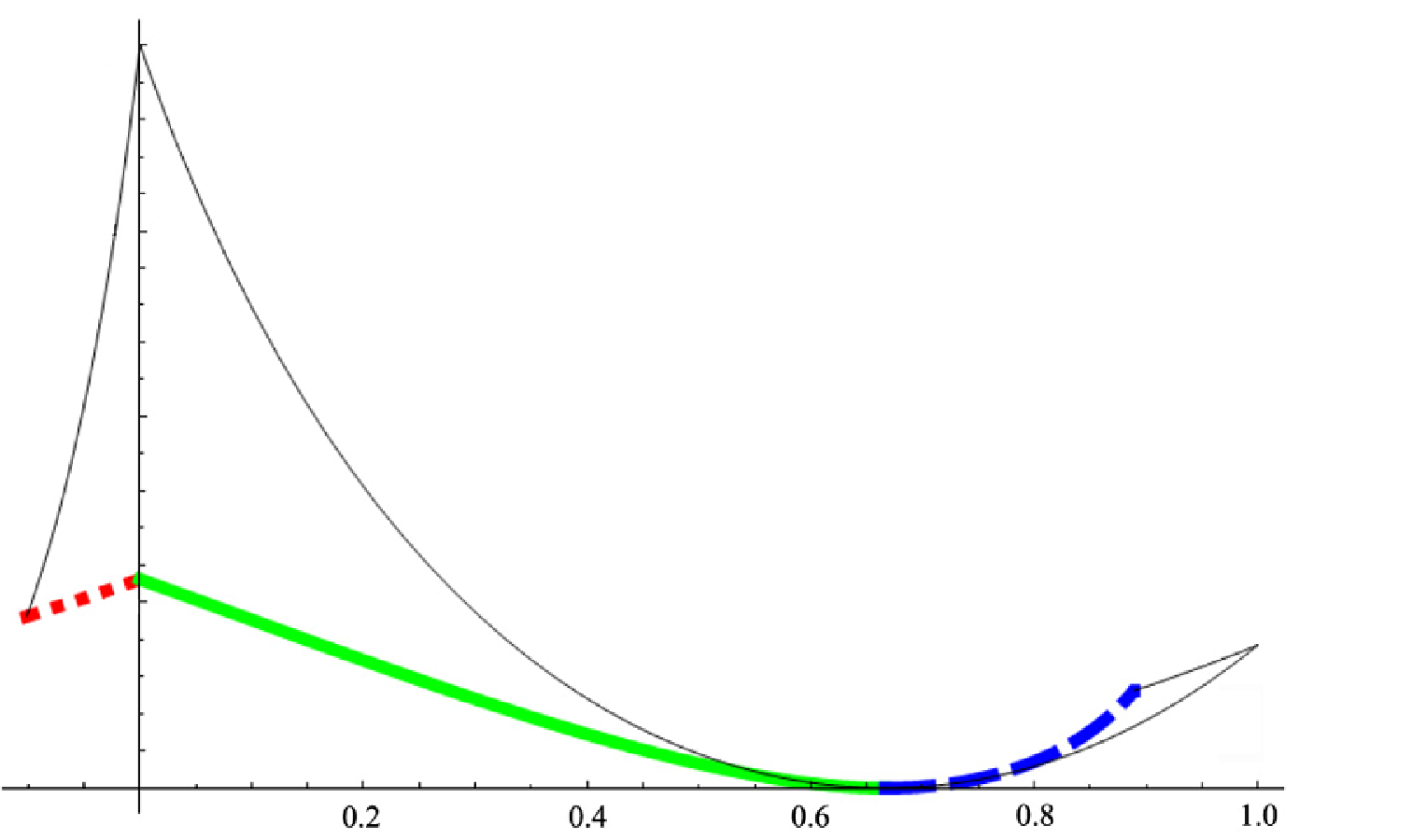}
\setlength{\unitlength}{1cm}
\put(-0.2,0.3){\fontsize{12}{14}\selectfont \makebox(0,0)[]{$\beta$ \strut}}
\put(-7.5,5){\fontsize{12}{14}\selectfont \makebox(0,0)[]{$C_m^2$ \strut}}
\put(-8.3,5){\fontsize{12}{14}\selectfont \makebox(0,0)[]{$\frac{4}{3}$ \strut}}
\put(-8.3,2.7){\fontsize{12}{14}\selectfont \makebox(0,0)[]{$\frac{2}{3}$ \strut}}
  \caption{(Color online) These curves show the amount of entanglement for the line state $\rho(\alpha,\frac{\beta}{2},\frac{\beta}{2})$,
  Eq.~(\ref{line}), in dependence of $\beta$ (compare with Fig.~\ref{P00P01P10}~(a)). The solid thin (black) curve is the amount of
  entanglement for all states lying on the boundary of positivity. The other graphs show the amount of entanglement for the boundary of CGLMP--Bell inequality violation. In detail, the
  dotted (red) graph illustrates the amount of entanglement for $\beta<0$. The solid
  thick (green) graph illustrates the amount of entanglement in the range $\beta=0$ to the separable state $\rho(\frac{1}{3},\frac{1}{3},\frac{1}{3})$ along the curved CGLMP boundary, while the dashed (blue) graph illustrates the region $\beta\geq\frac{2}{3}$.
   }
  \label{picturemeasure}
\end{figure}

Now, consider the class of line states with two equally weighted Bell states according to (\ref{eqmixed}). We evaluated the introduced bounds on a grid of $(\alpha,\beta)$ points with a step size of $0.02$ and found that they are exact (up to numerical precision) for this particular class of states. The numerical data can be represented with an accuracy of $10^{-6}$ by the expressions
\begin{align}
\label{amountent}
C_m^2(\rho_{\textrm{line}}(\alpha,\frac{\beta}{2},\frac{\beta}{2}))\;&=\frac{1}{27}\max\bigl\lbrace 0,8\alpha-\beta-2\bigr\rbrace^2  \ \   &\textrm{for}\;\alpha  \geq \frac{1}{4}+\frac{\beta}{8} \ ,\nonumber\\
C_m^2(\rho_{\textrm{line}}(\alpha,\frac{\beta}{2},\frac{\beta}{2}))\;&=\frac{2}{27}\max\bigl\lbrace
0,5\beta-4\alpha-2\bigr\rbrace^2  &\textrm{for}\;\alpha < \frac{1}{4}+\frac{\beta}{8} \ ,  \nonumber
\end{align}
for the $m$-concurrence $C_m^2$. The contour plot Fig.~\ref{measurecplot} shows that all states lying on lines parallel to the boundaries of separability contain the same amount of entanglement. In comparison to this, the strength of CGLMP-Bell inequality violation is not related to the distance to the separable states. Fig.~\ref{measurecplot} is an illustrative example that Bell inequality violations and the entanglement measures are non-monotonically related. Another example is illustrated in Fig.~\ref{picturemeasure} where
the amount of entanglement of states on the border of
CGLMP--Bell inequality violation ($\max I_3(\rho)=2$) and of positivity are plotted.
Both results demonstrate that there exists no monotone function relating the Bell inequality violation to the amount of entanglement.
For pure states this was already known, however, the surprising thing is that the same statement is true also for mixtures of states that all share the same amount of entanglement, i.e. $\{P_{k,l}\}$.

\section{Summary}
We presented the first detailed state space analysis of a Bell inequality in the context of geometry. The analysis necessitated to ascertain whether a Bell inequality is violated or obeyed for a given quantum state $\rho$. In order to solve this, we used a novel composite parameterization of the unitary group $\mathcal{U}$(d) and robust numerical methods (Sec.~\ref{nonlocalstates}). By means of this we achieved compliance with the orthonormality constraints and reduced the numerical search to $d^2-d$ real parameters for each involved observable. Our tests for the CGLMP--inequality (Sec.~\ref{sectionCGLMP}) showed that our method works within good numerical precision and that the used paramterization is beneficial with respect to numerical tractability. In Sec.~\ref{visCHSH} we began our geometric state space considerations concerning entanglement and nonlocality. We derived a regular tetrahedron spanned by the four Bell states for bipartite qubits. For this tetrahedron, the sets of entangled and CHSH--inequality violating states were determined with analytic methods. The graphical illustration of the tetrahedron (Fig.~\ref{tetranl}) provides a descriptive example that not all entangled states violate the CHSH--inequality. In Sec.~\ref{Wsimplex} we considered the state space geometry for the more general case of bipartite qudits. We focused on a generalization of the tetrahedron -- the so-called ``magic'' simplex $\mathcal{W}$ -- a class of states which is of special interest for quantum key distribution and entanglement distillation protocols. We gave a short review on the properties and the symmetries of $\mathcal{W}$ which are related to a classical discrete phase space (Fig.~\ref{ps}). In Sec.~\ref{visqutrits} we determined the nonlocal states of the simplex for the special case of bipartite qutrits ($d=3$) using the proposed numerical method. Here, we exploited a particular property of states containing uncolored noise (\ref{addnoise}) to further reduce the numerical effort. The boundaries of CGLMP--Bell inequality violation and separability were visualized for the class of the so-called ``line''
states and for the class of the so-called ``off-line'' states in Fig.~\ref{NL1} to Fig.~\ref{P00P01P10}. As could be seen in all cases, both boundaries are not simply related by a constant shift or scaling factor but rather are fundamentally different. In Sec.~\ref{measure} we considered the relation between entanglement measures and Bell inequality violations. We determined the exact value of the $m$-concurrence \cite{HHK1,HHmeasure} for a particular class of line states with two equally weighted parameters. Comparing the strength of CGLMP--Bell inequality violation with the
amount of entanglement, we geometrically demonstrated that both quantities are non-monotonically related.

In summary, we have seen that geometric considerations of the state space are an insightful way of studying the manifestations of entanglement. As we have shown, those approaches are suitable for revealing fundamental differences between entanglement and quantum nonlocality.\\

\textbf{Acknowledgements:} We thank T. Adaktylos, A.M. Adaktylos, A. Gabriel, F. Hipp, G. Krizek, H. Schimpf, M. Meyer
and H. Waldner for inspiring discussions and careful reading of the manuscript. C.S. and M.H. acknowledge financial support
from the Austrian FWF, Project P21947N16.\\

\end{document}